# On the solenoidal heat-flux in quasi-ballistic thermal conduction

Ashok T. Ramu[1†], and John E. Bowers[1]

[1] University of California, Santa Barbara, U.S.A.

[†] Corresponding author: ashok.ramu@gmail.com

Abstract

The Boltzmann transport equation for phonons is recast directly in terms of the heat-flux by means of iteration followed by truncation at the second order in the spherical harmonic expansion of the distribution function. This procedure displays the heat-flux in an explicitly coordinate-invariant form, and leads to a natural decomposition into two components, namely the solenoidal component in addition to the usual irrotational component. The solenoidal heat-flux is explicitly shown to arise by applying the heat-flux equation to a right-circular cylinder. These findings are important in the context of phonon resonators that utilize the strong quasi-ballistic thermal transport reported recently in silicon membranes at room temperature.

1. Introduction

With mounting evidence of significant room temperature ballistic phonon transport in silicon and other materials [1-3], several theories of non-Fourier heat transfer that account for this effect have arisen [4-8]. These theories solve the Boltzmann transport equation (BTE), usually in the isotropic and relaxation time approximations, either numerically or semi-numerically. Most of these are formulated as an equation for the phonon non-equilibrium distribution function, from which observables like the energy density and heat-flux may be extracted. However Ref. [9] reformulates the BTE directly in terms of the observables - heat-flux and temperature. This is achieved by means of a spherical harmonic expansion of the distribution function, followed by truncation at the next order (namely $l$=2) after the Fourier law.

Extension of these results to three spatial dimensions is far from trivial. Indeed, as we shall see in this paper, new physics materializes in the two- or three-dimensional (2D or 3D) heat-flux equations. This is because terms have to be considered involving the cross-derivatives in two or more spatial dimensions that are of course not present in a one-dimensional equation.

Any vector field can be decomposed into irrotational (curl-free, longitudinal) and solenoidal (divergence-free, transverse) components [10]. In an earlier article [11], we have discussed the quasi-ballistic irrotational heat-flux in detail in the context of frequency-domain thermoreflectance experiments. In this paper, for the first time we present a constitutive equation for the heat-flux that shows the existence of a solenoidal quasi-ballistic component. Furthermore, we show that the equations may be expressed in a manifestly coordinate-invariant form; thus even though the derivation will be in Cartesian coordinates, the result can be ported as-is to other coordinate systems, enhancing its utility in device thermal simulations in a variety of geometries.

As an illustrative example, we study 2D heat transfer in a right-circular cylinder. We will see that the heat-flux vector field has a non-zero curl inside the cylinder. This solenoidal heat-flux is seen to arise as the homogenous solution to a partial differential equation whose particular solution is the usual irrotational heat-flux. We comment upon the practical aspects of observation and application of the solenoidal heat-flux, and conclude by summarizing our findings.

## 2. The theory

Theoretical considerations begin with the time-independent Boltzmann transport equation in the isotropic and relaxation-time approximations. In what follows, symbols in bold font denote vectors, and those in regular font denote their magnitudes or other scalars. We assume the existence of a high-heat-capacity thermal reservoir with a well-defined, spatially varying temperature $T(\boldsymbol{r})$.

$$\boldsymbol{v}(\boldsymbol{k}) \cdot \nabla f = \frac{f_0 - f}{\tau(\boldsymbol{k})} \tag{1}$$

Here $\boldsymbol{v}(\boldsymbol{k})$ is the group velocity of a phonon mode of wave-vector $\boldsymbol{k}$, $f(\boldsymbol{r}, \boldsymbol{k})$ is the non-equilibrium distribution function, and $f_0(\boldsymbol{r}, \boldsymbol{k})$ is the equilibrium Bose distribution function corresponding to the reservoir temperature $T$. Under the assumption of isotropy, $\boldsymbol{v}(\boldsymbol{k})$ may be written as $v(k)\widehat{\boldsymbol{k}}$ where $\widehat{\boldsymbol{k}}$ is the unit vector in the direction of $\boldsymbol{k}$. Furthermore, the modal relaxation time $\tau(\boldsymbol{k}) = \tau(k)$ depends only upon the magnitude of the wave-vector. Writing the modal mean-free path $\Lambda(k) = v\tau$, we get

$$f = f_0 - \Lambda \widehat{\boldsymbol{k}} \cdot \nabla f \tag{2}$$

Iterating the BTE,

$$f = f_0 - \Lambda \widehat{\boldsymbol{k}} \cdot \nabla f_0 + \Lambda^2 (\widehat{\boldsymbol{k}} \cdot \nabla)(\widehat{\boldsymbol{k}} \cdot \nabla) f \tag{3}$$

Expanding $f$ in spherical harmonics $Y_{lm}(\theta, \phi)$ where $(\theta, \phi)$ are the polar angles of $\widehat{\boldsymbol{k}}$, and truncating at the *l*=1 term, we have

$$f = f_0 - \Lambda \widehat{\boldsymbol{k}} \cdot \nabla f_0 + \Lambda^2 (\widehat{\boldsymbol{k}} \cdot \nabla)(\widehat{\boldsymbol{k}} \cdot \nabla)(f_0 + \sum_{m=-1}^{+1} g_{1m} Y_{1m}(\theta, \phi)) \tag{4}$$

We multiply throughout by $\widehat{\boldsymbol{k}} v \hbar \omega(k)$ where $\omega(k)$ is the modal frequency, which again by the isotropic assumption depends only on the magnitude $k$. Then we integrate over the polar angles of $\widehat{\boldsymbol{k}}$. Recognizing that the LHS then becomes the modal heat-flux $\boldsymbol{q}(k, \boldsymbol{r})$, we have

$$\boldsymbol{q}(k,\boldsymbol{r}) = v\hbar\omega(k)\int d\Omega \widehat{\boldsymbol{k}} f_0 - \Lambda v\hbar\omega(k)\int d\Omega \widehat{\boldsymbol{k}}\widehat{\boldsymbol{k}}\cdot\nabla f_0 + \Lambda^2 v\hbar\omega(k)\int d\Omega \widehat{\boldsymbol{k}}(\widehat{\boldsymbol{k}}\cdot\nabla)(\widehat{\boldsymbol{k}}\cdot\nabla)f_0$$
$$+ \Lambda^2 v\hbar\omega(k)\int d\Omega \widehat{\boldsymbol{k}}(\widehat{\boldsymbol{k}}\cdot\nabla)(\widehat{\boldsymbol{k}}\cdot\nabla)\left(\sum_{m=-1}^{+1} g_{1m}Y_{1m}(\theta,\phi)\right)$$

(5)

Using the identities [8] $\int d\Omega \widehat{\boldsymbol{k}}(\widehat{\boldsymbol{k}}\cdot\nabla)^n f_0 = 0, n = 0,2,4\ldots$ and $\int d\Omega \widehat{\boldsymbol{k}}(\widehat{\boldsymbol{k}}\cdot\nabla)^n f_0 = \frac{4\pi}{n+2}\nabla^n f_0, n = 1,3,5\ldots$ the first and third terms on the RHS vanish, resulting in

$$\boldsymbol{q}(k,\boldsymbol{r}) = -\frac{4\pi}{3}\Lambda v\hbar\omega(k)\nabla f_0 + \Lambda^2 v\hbar\omega(k)\int d\Omega \widehat{\boldsymbol{k}}(\widehat{\boldsymbol{k}}\cdot\nabla)(\widehat{\boldsymbol{k}}\cdot\nabla)\left(\sum_{m=-1}^{+1} g_{1m}Y_{1m}(\theta,\phi)\right)$$

(6)

For further simplification, first we recognize that due to the orthogonality of spherical harmonics, the modal heat-flux may be written exactly in terms of the $g_{1m}$s alone:

$$\boldsymbol{q}(k,\boldsymbol{r}) = \int d\Omega \widehat{\boldsymbol{k}} v\hbar\omega(k) f = \frac{4\pi}{3} v\hbar\omega(k)(g^+\hat{e}_x + g^-\hat{e}_y + g^z\hat{e}_z) \quad (7)$$

Here $g^+ = \sqrt{3/8\pi}(-g_{11} + g_{1,-1})$; $g^- = -\sqrt{3/8\pi}i(g_{11} + g_{1,-1})$ and $g^z = \sqrt{3/4\pi}g_{10}$

Thus the heat-flux has been expressed in Cartesian coordinates, and the three quantities $g^+$, $g^-$ and $g^z$ are proportional to the Cartesian components of the heat-flux, $q_x(k,\boldsymbol{r})$, $q_y(k,\boldsymbol{r})$ and $q_z(k,\boldsymbol{r})$ respectively. Making two further observations; that $\sum_{m=-1}^{+1} g_{1m}Y_{1m}(\theta,\phi) = (g^+ sin\theta cos\phi + g^- sin\theta sin\phi + g^z cos\theta)$, and that the Bose distribution $f_0$ depends on the spatial coordinates through the temperature alone, we arrive from Eq. (6) at

$$\boldsymbol{q}(k,\boldsymbol{r}) = -\frac{4\pi}{3}\Lambda v\hbar\omega(k)\frac{\partial f_0}{\partial T}\nabla T$$
$$+ \Lambda^2 v\hbar\omega(k)\int d\Omega \widehat{\boldsymbol{k}}(\widehat{\boldsymbol{k}}\cdot\nabla)(\widehat{\boldsymbol{k}}\cdot\nabla)(g^+ sin\theta cos\phi + g^- sin\theta sin\phi + g^z cos\theta)$$

(8)

Replacing $\widehat{\boldsymbol{k}} = sin\theta cos\phi \hat{e}_x + sin\theta sin\phi \hat{e}_y + cos\theta \hat{e}_z$ everywhere on the second term of the RHS, we arrive at a set of 81 angular integrals of similar form. For example, the first integral is $\int d\Omega(\hat{e}_x sin\theta cos\phi)(\nabla_x sin\theta cos\phi)(\nabla_x sin\theta cos\phi)(g^+ sin\theta cos\phi) = (4\pi/3)0.6\hat{e}_x\nabla_x(\nabla_x g^+)$.

We choose to evaluate all the angular integrals numerically using the software MATLAB®, and use Eq. (7) to relate $g^+$ to $q_x(k,\boldsymbol{r})$, etc. Most of the 81 integrals evaluate to 0, and we are left with 21 terms. Rearranging, and with some effort (Appendix A), we arrive at the following equation for the modal heat-flux:

$$\boldsymbol{q}(k,\boldsymbol{r}) = -\frac{4\pi}{3}\Lambda v\hbar\omega(k)\frac{\partial f_0}{\partial T}\nabla T + \frac{3}{5}\Lambda^2\nabla(\nabla\cdot\boldsymbol{q}(k,\boldsymbol{r})) - \frac{1}{5}\Lambda^2\nabla\times(\nabla\times\boldsymbol{q}(k,\boldsymbol{r}))$$

(9)

Now we specialize to a "two-channel" model [12]. In order to derive the desired generalized form of the enhanced Fourier law[9] we suppose that apart from the thermal reservoir, there is a single dominant low-frequency (LF), low-heat-capacity phonon channel of constant mean-free path $\Lambda(k) = \Lambda^{LF}$. Applying the above equation to that channel, integrating over all wave-vector magnitudes $k$ comprising that channel, we have

$$\boldsymbol{q}^{LF}(\boldsymbol{r}) = -\kappa^{LF}\nabla T + \frac{3}{5}(\Lambda^{LF})^2\nabla(\nabla\cdot\boldsymbol{q}^{LF}(\boldsymbol{r})) - \frac{1}{5}(\Lambda^{LF})^2\nabla\times(\nabla\times\boldsymbol{q}^{LF}(\boldsymbol{r}))$$

(10)

Here $\kappa^{LF} = \frac{1}{3}\Lambda^{LF}vC^{LF}$, where $C^{LF} = \sum_k 4\pi\hbar\omega(k)\frac{\partial f_0}{\partial T}$ is the heat-capacity of the LF modes.

Combining with a Fourier law $\boldsymbol{q}^{HF}(\boldsymbol{r}) = -\kappa^{HF}\nabla T$ for the high-frequency reservoir, letting $\kappa^{LF} + \kappa^{HF} = \kappa$ (the net bulk thermal conductivity) and $\boldsymbol{q}^{LF}(\boldsymbol{r}) + \boldsymbol{q}^{HF}(\boldsymbol{r}) = \boldsymbol{q}$ (the net heat-flux), we arrive at the desired constitutive relation for the heat-flux in 3D quasi-ballistic transport:

$$\boldsymbol{q} = -\frac{1}{5}(\Lambda^{LF})^2\nabla\times(\nabla\times\boldsymbol{q}) + \frac{3}{5}(\Lambda^{LF})^2\nabla(\nabla\cdot\boldsymbol{q}) - \kappa\nabla T + \frac{3}{5}\kappa^{HF}(\Lambda^{LF})^2\nabla(\nabla^2 T)$$

(11)

Two key features of this equation become quite apparent: (a) the appearance of the subject of this article, the solenoidal term $-\frac{1}{5}(\Lambda^{LF})^2\nabla\times(\nabla\times\boldsymbol{q})$, and (b) the manifestly coordinate-invariant form of this equation, it being composed purely of invariant components.

In one dimension, the above equation reduces to the enhanced Fourier law of Ref. [9]. At first glance, it may seem hard to reconcile the fact that while in this work we truncate the spherical harmonic expansion at *l*=1 (see Eq. 4), we truncate it at *l*=2 in Ref. [9]. The equivalence is because in Eqs. (2) and (3), while iterating the BTE, we use the full distribution function, to all orders, on the RHS. Thus truncation at a lower order at the end of the iteration is sufficient.

We conclude this section by summarizing the assumptions under which the equations of this section are valid. We have linearized the Boltzmann transport equation using the relaxation-time approximation. We have assumed isotropy of the phonon dispersion relation and scattering rates. The constitutive relation for the heat-flux, Eq. (10) has been truncated at the leading order, namely $(\Lambda^{LF})^2$, in the phonon mean-free path. Beyond Eq. (9), the two-channel model for phonons has been applied, with the assumption of a single dominant long mean-free path mode. We discuss a sample application of the equations of this work in Sec. 4.

## 3. An example: Heat-flux in a right-circular cylinder

We further discuss the new term by considering phonon transport in a simple geometry. This example gives us insight into the boundary conditions (BCs) required to solve the new equation, which differ somewhat from the usual BCs for the Fourier law. Consider an infinitely long right-circular cylinder of radius $R$ in thermal steady-state, $\nabla \cdot \boldsymbol{q} = 0$, with temperature prescribed on the boundary as

$$T(R, \theta) = T_0 \sin\theta \tag{12}$$

Here $\theta$ is the angular coordinate. We further prescribe the heat-flux at the boundary as

$$\boldsymbol{q}(R, \theta) \cdot \hat{e}_r = -Q_0 \sin\theta \tag{13}$$

Here $\hat{e}_r$ is the usual radial unit-vector in the cylindrical-polar coordinate system. For later reference, we define $\hat{e}_\theta$ as the azimuthal unit vector of the cylindrical-polar system.

In order to bring out the essential physics of the solenoidal term, we consider only the leading order irrotational term, namely the Fourier-law term. Thus we consider the reduced equation

$$\boldsymbol{q} + \frac{1}{5}(\Lambda^{LF})^2 \nabla \times (\nabla \times \boldsymbol{q}) = -\kappa^{HF} \nabla T \tag{14}$$

This inhomogeneous differential equation has a particular solution,

$$\boldsymbol{q}_p = -\kappa^{HF} \nabla T \tag{15}$$

This may be verified by direct substitution. In order to solve for this heat-flux, we note from Eq. (14) that since $\nabla \cdot \boldsymbol{q} = 0$, $\nabla^2 T = 0$. This Laplace equation, together with the boundary condition Eq. (12), has the simple solution,

$$T(r, \theta) = T_0 \left(\frac{r}{R}\right) \sin\theta \tag{16}$$

This in turn, from Eq. (15), yields the particular solution, $\boldsymbol{q}_p$:

$$\boldsymbol{q}_p(r, \theta) = -\frac{\kappa^{HF} T_0}{R}(\sin\theta \hat{e}_r + \cos\theta \hat{e}_\theta) \tag{17}$$

This is in addition to the homogeneous solution, $\boldsymbol{q}_h$ which satisfies the equation

$$\boldsymbol{q}_h + \frac{1}{5}(\Lambda^{LF})^2 \nabla \times (\nabla \times \boldsymbol{q}_h) = 0 \tag{18}$$

By inspection, $\nabla \cdot \boldsymbol{q}_h = 0$, and with $\frac{1}{5}(\Lambda^{LF})^2 = 1/\epsilon^2$, we may write

$$\nabla^2 \boldsymbol{q}_h - \epsilon^2 \boldsymbol{q}_h = 0 \tag{19}$$

This is the vector Helmholtz equation in cylindrical-polar coordinates, with an imaginary eigenvalue. We follow the procedure of Gottlieb[13] to arrive at the following solution:

$$\boldsymbol{q}_h = (\alpha I_2(\epsilon r) + \beta I_0(\epsilon r))\sin\theta \hat{e}_r + (-\alpha I_2(\epsilon r) + \beta I_0(\epsilon r))\cos\theta \hat{e}_\theta \qquad (20)$$

$I_n$ is the modified Bessel function of the first kind of order $n$. Imposing the constraint $\nabla \cdot \boldsymbol{q}_h = 0$ we find that $\alpha = \beta$. The overall solution is the sum of the homogeneous and particular solutions:

$$\boldsymbol{q} = \boldsymbol{q}_h + \boldsymbol{q}_p = \left(\alpha(I_2(\epsilon r) + I_0(\epsilon r)) - \frac{\kappa^{HF}T_0}{R}\right)\sin\theta \hat{e}_r$$
$$+ \left(\alpha(-I_2(\epsilon r) + I_0(\epsilon r)) - \frac{\kappa^{HF}T_0}{R}\right)\cos\theta \hat{e}_\theta$$

(21)

Constant $\alpha$ may easily be determined from the heat-flux BC, Eq. (13), thus solving the problem. The curl of $\boldsymbol{q}$ may be determined analytically, using the properties of the modified Bessel functions[14]. Note that since there is no variation along the axis of the cylinder, the heat-flux lines are all perpendicular to it, and the curl of the heat-flux points along the cylindrical axis:

$$\nabla \times \boldsymbol{q} = -2\alpha\epsilon I_1(\epsilon r)\cos\theta \hat{e}_z \qquad (22)$$

We consider a silicon cylinder of radius $R$= 1 micron, with $T_0$, the amplitude of the temperature variation around the boundary set equal to 1 K, and with $Q_0$, the amplitude of the net heat-flux at the boundary set equal to $\frac{\kappa T_0}{R}$ where $\kappa$ = net thermal conductivity. We use parameters typical of thin silicon films[9]: $\kappa^{HF}$=30 W/m-K and $\kappa$=90 W/m-K. We find that in this example, the peak solenoidal heat-flux is an order of magnitude higher than the peak irrotational heat-flux, showing the importance of the solenoidal term. Fig. 1 shows a surface plot of the curl of the heat-flux (Eq. (22)) for two important cases: quasi-ballistic transport, with LF-mode mean-free path $\Lambda^{LF}$=0.6 micron and diffusive transport, with $\Lambda^{LF}$=0.15 micron. We see clearly that in the quasi-ballistic case, the curl is non-zero everywhere except close to the center of the cylinder axis. In the diffusive case however, the curl is close to zero everywhere except in a thin boundary layer of radial extent on the order of the mean-free path. This shows the necessity of phonon transport being in the quasi-ballistic regime in order to observe the solenoidal heat-flux described in this article.

Some comments are in order regarding the unorthodox boundary conditions required to render a solution to Eq. (14). Whereas with the ordinary Fourier law, the boundary temperature is necessary and sufficient, with our generalized Fourier law, both the temperature and the heat-flux are required at the boundary for a unique mathematical solution. Physically, this is because we are demanding more information from our equation than from the Fourier law – from the constitutive relations Eq. (11) or Eq. (14) can be derived both the LF- and reservoir- mode heat-fluxes. Thus we need to provide more information in the form of boundary conditions to seed the solutions. For instance, if a Joule heater is used to flux heat into the system under consideration, we will need to measure not only its temperature using some form of thermometry, but also record the "$I^2R$" heat supplied to it by means of an external circuit.

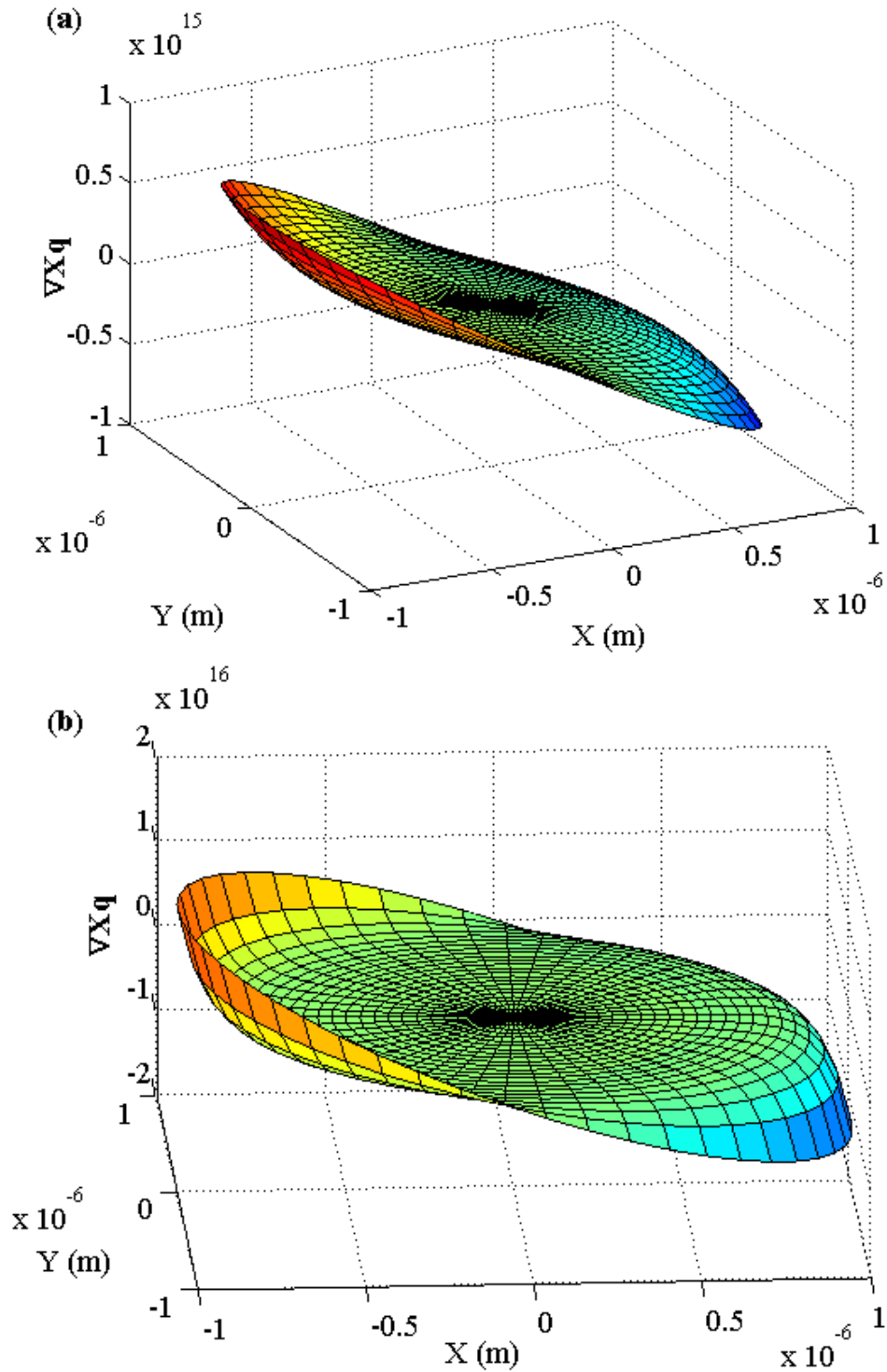

Fig. 1: Curl of the heat-flux (W/m³-K) vs. location relative to the axis of the cylinder. (a) MFP of the LF mode =0.6 micron (quasi-ballistic regime), and (b) MFP of the LF mode =0.15 micron (diffusive regime). Quasi-ballistic transport is essential to the observation of the solenoidal heat-flux inside the cylinder.

## 4. Practical considerations

We endeavor here to determine the conditions under which the solenoidal heat-flux may be observed in thermal systems of practical interest. An interesting application is suggested by the presence of the curl term in the constitutive relations, which renders possible a circulatory heat-flux.

But first, we address the question of consistency of a circulating heat-flux with the Second Law of Thermodynamics, one of whose statements is that heat cannot flow spontaneously from the cold side to the hot. To this end, we consider the following construction: consider a reversible heat-engine transferring heat from a hot reservoir to a cold reservoir and doing work in the process. We use this work to drive a reversible refrigerator that pumps heat from the cold reservoir to the hot. Thus we have established a circulating heat-flux in a cyclic system without doing external work.

A circulating heat-flux invites the promise of silicon phonon ring resonators that operate at least at a few tens of GHz, which is above what is typically attainable with thin-film bulk acoustic resonators. At such frequencies, the phonon wave-length is on the order of the lithographically attainable 1 micron, and the mean-free path is on the order of tens of microns. Thus the constructive interference and long lifetime conditions for a resonator may be met.

We now briefly examine this idea in the context of the equations developed in this work. In the previous section, we have shown in detail the existence of local curls in the solution to our equations. We now outline a solution where the lines of heat-flux close upon themselves. We consider for simplicity radially symmetric transport in an infinitely long circular cylinder. In the absence of temperature gradients, we have from Eq. (14) that:

$$\boldsymbol{q} + \frac{1}{5}(\Lambda^{LF})^2 \nabla \times (\nabla \times \boldsymbol{q}) = 0 \tag{23}$$

This equation has as a solution a radially symmetric vector field that clearly closes on itself:

$$\boldsymbol{q}(r,\theta) = C I_1(\epsilon r)\hat{e}_\theta \tag{24}$$

Here $\frac{1}{5}(\Lambda^{LF})^2 = 1/\epsilon^2$ and $I_1$ is the modified Bessel function of the first kind of order 0, as before. The solution may be verified by substitution in Eq. (23). $C$ is a constant to be determined from the boundary conditions. Thus we need to prescribe a tangential heat-flux as a boundary condition at some outer radius $r = R$ in Eq. (24). The problem of setting up a circulating phonon heat-flux at the boundary warrants further investigation. We note here that one may construct a circulating electronic (Peltier) heat-flux by passing a current through an electrically parallel combination of *n*- and *p*- type Peltier elements with identical resistances and Peltier coefficients. In this construction, despite there being no temperature gradients, a circulating heat-flux is maintained by doing external electrical work.

## 5. Conclusions

Generalization to three dimensions of an enhanced Fourier law yielded a manifestly coordinate-invariant constitutive equation for the heat-flux comprising a hitherto unrecognized solenoidal quasi-ballistic heat-flux term. Non-zero curl in quasi-ballistic transport was demonstrated in simulations of a cylinder; the quasi-ballistic transport regime was shown to be necessary for the observation of the solenoidal heat-flux component.

The finding of this article that heat-flux lines may indeed close upon themselves, coupled with the knowledge of strong quasi-ballistic effects in the micron-scale[1] in silicon at room-temperature, lead us to envisage silicon micro-ring phonon resonators that utilize the circulating solenoidal heat-flux for achieving resonance of thermal phonons. Coupling with optical modes may be predicted to lead to strong optical non-linear effects along the lines of Ref. [15].

Acknowledgments

We wish to thank Prof. Herbert Kroemer (University of California Santa Barbara, USA), Dr. R. S. Chandra (Siemens Corporate Research), Prof. Carl Meinhart (University of California Santa Barbara, USA), and Prof. Martin Kuball and Prof. Michael Uren (University of Bristol, UK) for helpful discussions. This work was funded by the National Science Foundation, USA under project number CMMI-1363207.

Appendix A:

We wish to establish that

$$\int (3/4\pi) d\Omega \hat{\boldsymbol{k}} (\hat{\boldsymbol{k}} \cdot \nabla)(\hat{\boldsymbol{k}} \cdot \nabla)(g^+ \sin\theta\cos\phi + g^- \sin\theta\sin\phi + g^z \cos\theta) = \frac{3}{5}\nabla(\nabla \cdot \boldsymbol{g}) - \frac{1}{5}\Lambda^2 \nabla \times (\nabla \times \boldsymbol{g}) \quad (A1)$$

where $\boldsymbol{g} = (g^+ \hat{e}_x + g^- \hat{e}_y + g^z \hat{e}_z)$

With $\hat{\boldsymbol{k}} = \sin\theta\cos\phi \hat{e}_x + \sin\theta\sin\phi \hat{e}_y + \cos\theta \hat{e}_z$ it may be seen that the LHS of Eq. (A1) is the sum of 81 integrals. Evaluating the angular integrals using the software MATLAB®, we find that only 21 are non-vanishing. Adding identical terms, this may further be condensed to 15 terms, listed below:

$$\int \left(\frac{3}{4\pi}\right) d\Omega \hat{\boldsymbol{k}}(\hat{\boldsymbol{k}} \cdot \nabla)(\hat{\boldsymbol{k}} \cdot \nabla)(g^+ \sin\theta\cos\phi + g^- \sin\theta\sin\phi + g^z \cos\theta) = 0.6\hat{e}_x \nabla_x \nabla_x g^+ + 0.4\hat{e}_y \nabla_y \nabla_x g^+ + 0.4\hat{e}_z \nabla_x \nabla_z g^+ + 0.2\hat{e}_x \nabla_y \nabla_x g^+ + 0.2\hat{e}_x \nabla_z \nabla_z g^+ + 0.4\hat{e}_x \nabla_x \nabla_y g^- + 0.2\hat{e}_y \nabla_x \nabla_x g^- + 0.6\hat{e}_y \nabla_y \nabla_y g^- + 0.4\hat{e}_z \nabla_y \nabla_z g^- + 0.2\hat{e}_y \nabla_z \nabla_z g^- + 0.4\hat{e}_x \nabla_x \nabla_z g^z + 0.2\hat{e}_z \nabla_x \nabla_x g^z + 0.4\hat{e}_y \nabla_z \nabla_y g^z + 0.2\hat{e}_z \nabla_y \nabla_y g^z + 0.6\hat{e}_z \nabla_z \nabla_z g^z$$

(A2)

We also have, from definitions,
$$\nabla(\nabla \cdot \boldsymbol{g}) = \hat{e}_x \nabla_x \nabla_x g^+ + \hat{e}_x \nabla_x \nabla_y g^- + \hat{e}_x \nabla_x \nabla_z g^z + \hat{e}_y \nabla_y \nabla_x g^+ + \hat{e}_y \nabla_y \nabla_y g^- + \hat{e}_y \nabla_y \nabla_z g^z + \hat{e}_z \nabla_z \nabla_x g^+ + \hat{e}_z \nabla_z \nabla_y g^- + \hat{e}_z \nabla_z \nabla_z g^z \quad (A3)$$

and,
$$\nabla^2 \boldsymbol{g} = \hat{e}_x \nabla_x \nabla_x g^+ + \hat{e}_x \nabla_y \nabla_y g^+ + \hat{e}_x \nabla_z \nabla_z g^+ + \hat{e}_y \nabla_x \nabla_x g^- + \hat{e}_y \nabla_y \nabla_y g^- + \hat{e}_y \nabla_z \nabla_z g^- + \hat{e}_z \nabla_x \nabla_x g^z + \hat{e}_z \nabla_y \nabla_y g^z + \hat{e}_z \nabla_z \nabla_z g^z \quad (A4)$$

It can be seen by comparing terms that
$$\int (3/4\pi) d\Omega \hat{\boldsymbol{k}}(\hat{\boldsymbol{k}} \cdot \nabla)(\hat{\boldsymbol{k}} \cdot \nabla)(g^+ \sin\theta\cos\phi + g^- \sin\theta\sin\phi + g^z \cos\theta) = = 0.4\nabla(\nabla \cdot \boldsymbol{g}) + 0.2\nabla^2 \boldsymbol{g} \quad (A5)$$

whence we arrive at Eq. (A1) by noting the identity $\nabla \times (\nabla \times \boldsymbol{g}) = \nabla(\nabla \cdot \boldsymbol{g}) - \nabla^2 \boldsymbol{g}$